\documentclass[sort&compress,12pt]{article}%
\usepackage{amssymb}
\usepackage{amsfonts}
\usepackage{amsmath}
\usepackage{breqn}
\usepackage{graphicx}
\usepackage[T1]{fontenc}
\usepackage[utf8]{inputenc}
\usepackage{xcolor}

\begin{document}

\title{The Noncommutative Coulomb Potential}
\author{B.C. Wang$^{a}$, E.C. Brenag$^{b}$ R.G.G. Amorim$^{a,c,d}$,\\ V.C. Rispoli$^{a}$, S.C. Ulhoa$^{c,d}$\\${}^{a}$ Universidade de Bras\'{\i}lia, Faculdade Gama,\\72444-240, Bras\'{\i}lia, DF, Brazil\\${}^{b}$ Universidade de Bras\'{\i}lia, Faculdade de Tecnologia,\\70910-900, Bras\'{\i}lia, DF, Brazil\\${}^{c}$International Centre for Condensed Matter Physics$,$\\Instituto de F\'{\i}sica, Universidade de Bras\'{\i}lia,\\70910-900, Bras\'{\i}lia, DF, Brazil\\${}^{d}$Canadian Quantum Research Center,\\
204-3002 32 Ave Vernon, BC V1T 2L7  Canada\\}
\maketitle

\begin{abstract}
In this work, we analyze the noncommutative three-dimensional Coulomb potential problem. For this purpose, we used the Kustaanheimo-Stiefel mapping to write the Schrödinger equation for Coulomb potential in a solvable way. Then,  the noncommutative hydrogen-like atoms were treated, and their energy levels were found. In addition, we estimate a bound for the noncommutativity parameter.

%Pacs: 03.65.Ca; 03.65.Db; 11.10.Nx

\end{abstract}

\section{Introduction}
Quantum mechanics was born out of experimental observation \cite{qm1,qm2}.  In this context, the measurement of the hydrogen spectrum was fundamental for the establishment of atomic theory \cite{hydro3,hidro1,hidro2}.  Even today, several measurements continue to be made in the hope of verifying both the limit of application of the usual quantum mechanics and the veracity of some unorthodox ideas.  One of these ideas stands out and will be the main object of this article, the noncommutativity of space.  Certainly, noncommutativity per se is as old as quantum mechanics, given that the main characteristic of it is a noncommutation relationship between coordinate and moment, which takes place in the phase space.  On the other hand, the extension of this concept to the coordinate space is not widespread.  To be more precise, this concept was introduced by Snyder in 1947~\cite{jackiw,snyder1,snyder2}, this proposition established an uncertainty principle for space, that is, it would not be possible to measure two coordinates of space at the same time up to a certain scale.  This is translated into the relationship
 $$
 [\widehat{x}^{\mu}, \widehat{x}^{\nu}] = i \Theta^ {\mu \nu} \,,
 $$
 where $ \Theta $ is the noncommutative space parameter.  This inaugurated the so-called noncommutative geometry. The noncommutative parameter marks the scale at which such effects are relevant. Likewise, as with usual quantum mechanics, the Planck constant establishes the scale at which quantum effects are important \cite{snyder2}. Operators in this noncommutative geometry are usually built using the product of Moyal or star product, such product is defined as \cite{snyder1}
\begin{equation}
f(x) \star g(x) \equiv \exp
\left(  {i \over 2} \Theta^{\mu\nu}  {\partial \over \partial x^\mu}
{\partial \over \partial y^\nu} \right) f(x) g(y) |_{y \rightarrow x}, 
\end{equation}
so an operator constructed from the spatial coordinate is
$$\widehat{x}^{\mu} = x^\mu\star\,.$$
Thus, a Lagrangian density can be rewritten in noncommutative terms by replacing the ordinary product with the star product.  It is important to note that there are several representations for these operators as long as the commutation relation is maintained.

The applications of noncommutative geometry are numerous, it is possible to highlight some approaches to quantum gravitation~\cite{carga} and the formulation of quantum mechanics in phase space~\cite{phase}.  Particularly the spectrum of the hydrogen atom may still reveal a possible non-commutative nature of space.  Therefore, it is possible to associate the limit of the noncommutative parameter to the experimental error of the measurement.  This approach is in fact a speculation that requires the development of increasingly accurate techniques to be refuted.  In this article, we will use a formulation of non-commutative quantum mechanics in the phase space. Thus the error in measuring the transition frequency from 1S to 2S hydrogen states is associated with the non-commutative parameter. In fact, in a previous article, this problem was addressed for the hydrogen atom in two dimensions.  But the model was not very realistic as it could not handle a transition between levels 1S and 2S.  So the two-dimensional hydrogen atom works just like a toy model.

This paper is organized as follows. In section 2, we present the Schrödinger equation for Coulomb potential. The Kustaanheimo-Stiefel transformation is presented in section 3. In section 4, we analyze the noncommutative Coulomb potential. The correction of the energy levels and an estimative of the noncommutative parameter is presented in section 5.  Finally,
in section 6, we present our concluding remarks.

\section{Schrödinger Equation for Coulomb Potential}

The Coulomb potential is a fundamental tool in quantum mechanics because several systems can be simulated by this singular potential. For instance, the hydrogen-like atoms are studied in this perspective. Then, in this section we discuss about the Schrödinger equation for hydrogen-like atoms. In this sense, the Hamiltonian for those systems can be written by
\begin{equation}\label{h1}
H=\frac{1}{2m}(p_{x}^2+p_{y}^2+p_{z}^2)+\frac{k}{(x^2+y^2+z^2)^{1/2}},
\end{equation}
where $p_i=(p_{x}, p_y, p_z)$ stands for the momentum of electron in directions $x$, $y$ and $z$ respectively; $x_i=(x, y, z)$ represents the electron coordinates and the constant $k=-\frac{Ze^{2}}{4\pi\epsilon_o}$, where $e$ is the elemental electrical charge, $Z$ is the atomic number and $\epsilon_o$ is the vacuum electrical permissivity. For quantize the Hamiltonian given in Eq.(\ref{h1}), as usual, the momentum operators are given by $\widehat{p}_{x}=-i\hbar\frac{\partial}{\partial x}$, $\widehat{p}_{y}=-i\hbar\frac{\partial}{\partial y}$ and $\widehat{p}_{z}=-i\hbar\frac{\partial}{\partial z}$, where $\hbar=h/2\pi$ and $h$ is the Planck constant. This operators can be written in a compacted presentation as $\widehat{p}_i=-i\hbar\frac{\partial}{\partial x_i}$. In this context, the Heisenberg commutation relations are satisfied, 
\begin{equation}\nonumber
[\widehat{x}_i,\widehat{p}_j]=i\delta_{ij}\hbar,
\end{equation}
where $\widehat{x}_i=(\widehat{x},\widehat{y},\widehat{z})$ represents the coordinate  and $\delta_{ij}$ is the Kronecker's delta symbol.

Our aim in this work is analyze the non-commutative Coulomb potential. For this propose we define the coordinates operators as
\begin{equation}\label{nc1}
\widehat{x}_i=x_i+i\frac{\theta_{ij}}{2}\frac{\partial}{\partial x_j},
\end{equation}
where $\theta_{ij}$ is an anti-symmetric tensor called non-commutativity parameter. Notice that repeated index indicate sum.  In this sense, the coordinate operators satisfies
\begin{equation}\label{nc2}
[\widehat{x}_i,\widehat{x}_j]=i\theta_{ij}.
\end{equation}

However, the treatment of Hamiltonian given in Eq.(\ref{h1}), when we replace the variables by the operators, is difficult because the operators in the denominator of the potential energy term. For this reason, in the next section we present a transformation that put the system in a more suitable way. 

\section{ Kustaanheimo-Stiefel Transformation}

The Kustaanheimo-Stiefel transformation appears in an attempt to generalize
characteristics of the two-dimensional Levi-Civita transformation
to dimensions $d>2$ \cite{ks,kliber}. Due to a Hurwitz theorem \cite{hurwitz}, it is
not possible to construct such a transformation for the three-dimensional
case, being the four-dimensional one the close as possible to have
an extension \cite{ks,kliber}. This transformation is widely used in the literature
to show that the quantization of the hydrogen-like atom Coulomb potential problem is equivalent
to the quantization of a four-dimensional isotropic harmonic oscillator
supplemented with a constraint. Then, Kustaanheimo-Stiefel transformation
is defined as a $\mathbb{R}^{4}\rightarrow\mathbb{R}^{3}$ surjective
mapping satisfying the following set of equations
\begin{eqnarray}
x & =&2\left(ut-vw\right)\nonumber\\
y & =&2\left(uw+vt\right)\nonumber\\
z & =&u^{2}+v^{2}-t^{2}-w^{2}\label{hh1}
\end{eqnarray}
with the additional condition
\begin{equation}
vdu-udv-wdt+tdw=0.\label{constraint}
\end{equation}

Given the Eqs.(\ref{hh1}) it is straightforward to show that
\begin{eqnarray}
\frac{\partial}{\partial x} & =&\frac{1}{2\rho}\left(t\frac{\partial}{\partial u}-w\frac{\partial}{\partial v}+u\frac{\partial}{\partial t}-v\frac{\partial}{\partial w}\right)\nonumber\\
\frac{\partial}{\partial y} & =&\frac{1}{2\rho}\left(w\frac{\partial}{\partial u}+t\frac{\partial}{\partial v}+v\frac{\partial}{\partial t}+u\frac{\partial}{\partial w}\right)\nonumber\\
\frac{\partial}{\partial z} & =&\frac{1}{2\rho}\left(u\frac{\partial}{\partial u}+v\frac{\partial}{\partial v}-t\frac{\partial}{\partial t}-w\frac{\partial}{\partial w}\right) \label{hh3}
\end{eqnarray}
where $\rho=\sqrt{x^{2}+y^{2}+z^{2}}=u^{2}+v^{2}+t^{2}+w^{2}$. Then, by Eqs.(\ref{hh3}),
the momentum operators can be written as follows
\begin{eqnarray}
p_{x} & =&\frac{1}{2\rho}\left(tp_{u}-wp_{v}+up_{t}-vp_{w}\right)\nonumber\\
p_{y} & =&\frac{1}{2\rho}\left(wp_{u}+tp_{v}+vp_{t}+up_{w}\right)\nonumber\\
p_{z} & =&\frac{1}{2\rho}\left(up_{u}+vp_{v}-tp_{t}-wp_{w}\right).\label{hh2}
\end{eqnarray}

Applying Eqs.(\ref{hh2}) in Eq.(\ref{h1}),
we obtain the following transformed Hamiltonian 
\begin{equation}
H=\frac{1}{2m\rho}\left(p_{u}^{2}+p_{v}^{2}+p_{t}^{2}+p_{w}^{2}\right)+\frac{2k}{\rho}.\label{h5}
\end{equation}
Then, the hypersurface defined by $H=E$ is given in this new coordinate system by 
\begin{equation}
\frac{1}{2m}\left(p_{u}^{2}+p_{v}^{2}+p_{t}^{2}+p_{w}^{2}\right)-E(u^{2}+v^{2}+t^{2}+w^{2})=-2k.\label{a2}
\end{equation}
Finally, since Kustaanheimo-Stiefel transformation is nonbijective, an additional constraint is necessary to ensure a univalued wavefunction when the problem is quantized and the wavefunction is applied to the operators \cite{kliber,campos}. Thus, based on the Eq.(\ref{constraint}), the additional constraint is given by
\begin{equation}\label{a2a}
vp_u-up_v-wp_t+tp_w=0.
\end{equation}
It is important to note that, due to the energy level analysis that is performed in this paper, Eq.\eqref{a2a} will not be relevant and then will not be used in the following sections.
%Equation (\ref{a2}) is the main result of this section and is the
%one to be used from now on. 

\section{Analysis of Non-Commutative Coulomb Potential}
%Using the Kustaanheimo-Stiefel Mapping presented in the previous section, the Hamiltonian given in Eq.(\ref{h1}) can be written as
%\begin{equation}\label{a1}
%H=\frac{1}{2m\rho}(p_u^2+p_v^2+p_t^2+p_w^2)+2\frac{k}{\rho},
%\end{equation}
%Taking an hypersurface of energy such as $H=E$, we obtain
%\begin{equation}
%\frac{1}{2m}(p_u^2+p_v^2+p_t^2+p_w^2)-E(u^2+v^2+t^2+w^2)=-2k.
%\end{equation}
In this paper, we will analyze this problem in two different non-commutative plans, $[\widehat{u},\widehat{v}]=i\alpha$ and $[\widehat{t},\widehat{w}]=i\beta$, where $\alpha$ and $\beta$ are the non-commutative parameters in $u-v$ and $t-w$ planes, respectively.  In this way, the coordinate operators are
\begin{eqnarray}\label{a3}
\widehat{u}=u+i\frac{\alpha}{2}\frac{\partial}{\partial v}, && \widehat{v}=v-i\frac{\alpha}{2}\frac{\partial}{\partial u}, \nonumber\\
\widehat{t}=t+i\frac{\beta}{2}\frac{\partial}{\partial w}, && \widehat{w} = w-i\frac{\beta}{2}\frac{\partial}{\partial t}.
\end{eqnarray}
The momentum operators are given by
\begin{eqnarray}\label{a4}
\widehat{p}_u=-i\hbar\frac{\partial}{\partial u}, &&
\widehat{p}_v=-i\hbar\frac{\partial}{\partial v},\nonumber\\
\widehat{p}_t=-i\hbar\frac{\partial}{\partial t}, &&
\widehat{p}_w=-i\hbar\frac{\partial}{\partial w}.
\end{eqnarray}
Substituting the operators given by Eqs.\eqref{a3}--\eqref{a4} in Eq.(\ref{a2}) and also applying the wavefunction $\psi\equiv\psi(u,v,t,w)$, we get
\begin{eqnarray}\label{a5}
&&-\left(\frac{\hbar^2}{2m}-\frac{E\alpha^2}{4}\right)\left(\frac{\partial^2}{\partial u^2}+\frac{\partial^2}{\partial v^2}\right)\psi-i\alpha E\left(u\frac{\partial}{\partial v}-v\frac{\partial}{\partial u}\right)\psi-E(u^2+v^2)\psi\nonumber\\
&&-\left(\frac{\hbar^2}{2m}-\frac{E\beta^2}{4}\right)\left(\frac{\partial^2}{\partial t^2}+\frac{\partial^2}{\partial w^2}\right)\psi-i\beta E\left(t\frac{\partial}{\partial w}-w\frac{\partial}{\partial t}\right)\psi-E(t^2+w^2)\psi\nonumber\\
&&=-2k\psi.
\end{eqnarray}
Assuming that 
$\psi(u,v,t,w)=f(u,v)g(t,w)$, we obtain the following set of equations
\begin{eqnarray}\label{a6}
&-&\left(\frac{\hbar^2}{2m}-\frac{E\alpha^2}{4}\right)\left(\frac{\partial^2f}{\partial u^2}+\frac{\partial^2f}{\partial v^2}\right)-i\alpha E\left(u\frac{\partial f}{\partial v}-v\frac{ f}{\partial u}\right)\\&-&E(u^2+v^2)f+(k-\kappa_1)f=0\nonumber,
\end{eqnarray}
and
\begin{eqnarray}\label{a7}
&-&\left(\frac{\hbar^2}{2m}-\frac{E\beta^2}{4}\right)\left(\frac{\partial^2g}{\partial t^2}+\frac{\partial^2g}{\partial w^2}\right)-i\beta E\left(t\frac{\partial g}{\partial w}-w\frac{ \partial g}{\partial t}\right)\\&-&E(t^2+w^2)g+(k-\kappa_2)g=0,\nonumber
\end{eqnarray}
where $\kappa_1$ and $\kappa_2$ are separation constants that satisfies $\kappa_1=-\kappa_2=\mathrm{constant}$. Equations (\ref{a6}) and (\ref{a7}) are identical, except for a few constants. So, we will solve only one of them. In this way, lets consider Eq.(\ref{a6}). Performing the change of variables $q=u^2+v^2$, we immediately obtain
\begin{equation}\label{a8'}
q\frac{d^2f(q)}{dq^2}+\frac{df(q)}{dq}-\frac{1}{\lambda}[Eq-(k-\kappa_1)]f(q)=0,
\end{equation}
where $\lambda=-4\left(\frac{\hbar^2}{2m}-\frac{E\alpha^2}{4}\right)$.

Considering that $f(q)=e^{-\gamma q}\phi(q)$, where $\gamma=\sqrt{\frac{E}{\lambda}}$, we can write the Eq.\eqref{a8'} as
\begin{equation}\label{a8}
q\frac{d^2\phi(q)}{dq^2}+(1-2\gamma q)\frac{d\phi(q)}{dq}-\left[\gamma-\frac{(k-\kappa_1)}{\lambda}\right]\phi(q)=0.
\end{equation}

Finally, performing the change of variable $w=2\gamma q$, we obtain
\begin{equation}\label{a9}
w\frac{d^2\phi(w)}{dw^2}+(1-w)\frac{d\phi(w)}{dw}+\frac{1}{2\gamma}\left(\frac{(k-\kappa_1)}{\lambda}-\gamma\right)\phi(w)=0.
\end{equation}
Note that Eq.(\ref{a9}) has the following form
\begin{equation}
w\phi''+(1-w)\phi'+l\phi=0,
\end{equation}
which is the Laguerre differential equation. If $l$ is an integer $l=0,1,2,3,\ldots$ the solution of Laguerre’s equation is given by Laguerre polynomials 
\begin{equation}
L_l(x)=\frac{e^x}{l!}\frac{d^l}{dx^l}\left(e^{-x}x^l\right),
\end{equation}
as given by the Rodrigues' formula \cite{rodrigues}. In this sense, we obtain the solution for Eq.\eqref{a6}
\begin{equation}\label{a10}
f(u,v)=e^{-\gamma(u^2+v^2)}L_{l_1}(2\gamma(u^2+v^2)),
\end{equation}
where $l_1=0,1,2,3,\ldots$. In a totally analogous way, the solution of Eq.(\ref{a7}) can be written as
\begin{equation}\label{a10}
g(t,w)=e^{-\kappa(t^2+w^2)}L_{l_2}(2\kappa(t^2+w^2)),
\end{equation}
where $\kappa=\sqrt{\frac{E}{\eta}}$, $\eta=-4\left(\frac{\hbar^2}{2m}-\frac{E\beta^2}{4}\right)$ and $l_2=0,1,2,3,\ldots$. Thus, the solution of Eq.(\ref{a5}) is
\begin{equation}\label{a11}
\psi(u,v,t,w)=e^{-[\gamma(u^2+v^2)+\kappa(t^2+w^2)]}L_{l_1}(2\gamma(u^2+v^2))L_{l_2}(2\kappa(t^2+w^2)).
\end{equation}
Notice that the eigenvalues relative of this functions are degenerate. Eq.(\ref{a2a}) is useful to fix $l_1$ and $l_2$, however, we follow for other way. The degree of degeneracy is $l+1$, where $l=l_1+l_2$. Only the ground state $l_1=l_2=0$ is non-degenerate. In the next section we use the results of this section to determine the energy levels for the considered system.
\section{The correction of energy levels and an estimative for non-commutative parameter}

From Laguerre's differential equation, Eq.\eqref{a9}, we can write
\begin{equation}\label{a12}
l_1=\frac{1}{2}\left(\frac{(k-\kappa_1)}{\lambda\gamma}-1\right),
\end{equation}
\begin{equation}\label{a13}
l_2=\frac{1}{2}\left(\frac{(k-\kappa_2)}{\eta\kappa}-1\right).
\end{equation}
Equations \eqref{a12} and \eqref{a13} can be arranged as
\begin{equation}\label{a14}
2(l_1+l_2+1)=\frac{(k-\kappa_1)}{\gamma\lambda}+\frac{(k-\kappa_2)}{\eta\kappa}.
\end{equation}
For a suitable analysis, we consider the particular case where $\beta=\alpha$.  Using the fact that separation constants satisfies $\kappa_1+\kappa_2=0$, we get
\begin{equation}\label{a15}
l_1+l_2+1=\frac{k}{\sqrt{\lambda E}}.
\end{equation}
In this context we can determine the energy levels $E$ from the solution of the following equation
\begin{equation}\label{a16}
E^2-\frac{2\hbar^2}{m\alpha^2}E-\frac{k^2}{(l_1+l_2+1)^2\alpha^2}=0.
\end{equation}
If we define $n=l_1+l_2+1$, solutions of Eq.(\ref{a16}) take the form
\begin{equation}\label{a17}
E_n=\frac{\hbar^2}{m\alpha^2}\pm\frac{\hbar^2}{m\alpha^2} \sqrt{1+\frac{m^2k^2\alpha^2}{\hbar^4n^2}}.
\end{equation}
If $\frac{m^2k^2\alpha^2}{\hbar^4n^2} \ll 1$ the use of binomial series $(1+x)^j = 1+jx+j(j-1)x^2/2+\ldots$, up to the second order term, gives 
\begin{equation}\label{a1}
E_n\approx -\frac{me^4}{32\pi^2\epsilon_o^2\hbar^2n^2}+\frac{m^3e^8\alpha^2}{2048\pi^4\epsilon_o^4 \hbar^6 n^4}.
\end{equation}
In the limit $\alpha \rightarrow 0$, the result of the usual spectral for hydrogen atom is obtained. In addition, the first order term in $\alpha$ do not contribute to the energy of this system.

In this sense, the non-commutative correction, $\Delta E_{NC}$, for the energy is given by
\begin{equation}\label{h15}
\Delta E_{NC}\approx\frac{m^3e^8\alpha^2}{2048\pi^4\epsilon_o^4 \hbar^6 n^4}.
\end{equation}

We can use the experimental value for $1S\rightarrow 2S$ frequency transition in the hydrogen atom is $\nu_{1S\rightarrow 2S}=(2 466 061 102 474 851 \pm 34) Hz$ \cite{exp} and Eq.(\ref{h15}) to estimate the upper bound on the non-commutativity parameter $\alpha$. Then, the theoretical value for the error in transition $1S \rightarrow 2S$, denoted by $\Delta E_{1\rightarrow 2}$, is given by
\begin{equation}\nonumber
\Delta E_{1\rightarrow 2}\approx \frac{m^3e^8\alpha^2}{2048\pi^4\epsilon_{o}^4\hbar^6}\left[\frac{15}{16}\right],
\end{equation}

For the three-dimensional case, the energy is given by $\Delta E_{1\rightarrow 2}= h\Delta\nu$, where $h$ is Planck constant. So, we have the following bound
\begin{equation}\nonumber
\alpha\lesssim\left[\frac{16}{15}\frac{4096\pi^5\epsilon_{o}^4 \hbar^7\Delta \nu}{m^3e^8}\right]^{1/2}.
\end{equation}
Performing the calculations, we obtain $\alpha \lesssim 1.12 \cdot 10^{-17} m$.

\section{Concluding remarks}
Using the Kustaanheimo-Stiefel transformation, we treated the non-trivial problem of the non-commutative Coulomb potential. As a result, we obtain the solution of the Schrödinger equation for this system and calculate the energy levels. Using the spectrum obtained and experimental data, we estimated the non-commutativity parameter $\theta$, which has order of magnitude of $10^{-34} m^2$, and the non-commutative effects will be relevant to a length smaller than $10^{-17} m$. These results are in agreement with the literature \cite{exp, rodrigues}, bearing in mind that there is no prediction of energy dependence with a non-commutativity parameter linearly. In a different perspective to the two-dimensional case, in this model the correct value to usual energy levels is obtained when the non-commutative parameter vanishes to zero, i.e, $\alpha\rightarrow 0$.  In addition, this model does allow to verify a transition between
states $1s$ and $2s$, because $n$, which appears in the energy, is an any integer bigger than zero.

It is worth mentioning that the association of the non-commutative parameter with the experimental error in measuring the transition frequency of the hydrogen atom shells is legitimate. The main reason for this is that below the experimental limit there is room for speculation. We hope that more and more precise measures can reach the predicted  limit of non-commutativity of space and that the hydrogen atom will be used, once again, as a support point for fundamental discoveries of nature.  

\section*{Acknowledgements}

This work was partially supported by CNPq of Brazil.

\end{document}